\documentclass[
	11pt,
	a4paper,
	journal,
	compsoc
]{IEEEtran}

\usepackage{cite}
\usepackage{amsmath,amssymb,amsfonts}
\usepackage{algorithmic}
\usepackage{graphicx}
\usepackage{textcomp}
\usepackage{xcolor}
\usepackage{xspace}
\usepackage{listings}
\usepackage{hyperref}
\usepackage{booktabs}
\usepackage{orcidlink}

\newcommand{\kpcbn}{\texttt{kicad-pycbnew}\xspace}
\newcommand{\kicadpcb}{\texttt{.kicad\_pcb}\xspace}
\newcommand{\pcbnew}{\texttt{pcbnew}\xspace}
\newcommand{\usc}{\_\allowbreak{}}
\newcommand{\fnametab}[1]{\texttt{\footnotesize #1}}

\definecolor{codegreen}{rgb}{0,0.6,0}
\definecolor{codegray}{rgb}{0.5,0.5,0.5}
\definecolor{codepurple}{rgb}{0.58,0,0.82}
\definecolor{backcolour}{rgb}{0.95,0.95,0.92}

\lstdefinestyle{mystyle}{
    backgroundcolor=\color{backcolour},
    commentstyle=\color{codegreen},
    keywordstyle=\color{magenta},
    numberstyle=\tiny\color{codegray},
    stringstyle=\color{codepurple},
    basicstyle=\ttfamily\footnotesize,
    breakatwhitespace=false,
    breaklines=true,
    captionpos=b,
    keepspaces=true,
    numbers=left,
    numbersep=5pt,
    showspaces=false,
    showstringspaces=false,
    showtabs=false,
    tabsize=2,
    frame=single,
    escapeinside={(*@}{@*)}
}
\definecolor{mygray}{gray}{0.6}

\title{\textbf{\texttt{kicad-pycbnew}}: A Python Framework for Automated, High-Precision PCB Design in Particle Detector Instrumentation
}

\author{\orcidlinki{Vincent \textsc{Raspal}}{0009-0008-7997-0152}\\
\IEEEauthorblockA{\textit{Université Clermont Auvergne, CNRS IN2P3, LPCA}\\
F-63000 Clermont-Ferrand, France \\
\href{mailto:vincent.raspal@clermont.in2p3.fr}{\texttt{\small vincent.raspal@clermont.in2p3.fr}}}
}

\begin{document}
\maketitle

\begin{abstract}
\kpcbn is an open-source Python library for the automated, script-based generation of KiCad PCB designs, tailored to the precise geometric and impedance constraints of particle detector readout electronics.
By enabling the programmatic definition of copper traces, zones, vias, and footprints ---~without a graphical interface~--- it addresses the challenges of designing large-scale, repetitive, or parameterized layouts (e.g., Resistive Plate Chamber (RPC) strips or controlled-impedance signal paths).
Key features include constraint-driven routing, parallel track arrays with rounded corners, via grids with exclusion zones, and full compatibility with KiCad’s native file format.
Standalone operation (no KiCad runtime required) facilitates integration into CI/CD pipelines and ensures reproducibility.
The library, released under the MIT License, includes validated examples such as trapezoidal RPC readout boards and extension boards for the COMET experiment, alongside comprehensive tests and documentation.
\end{abstract}


\section{Introduction}
\IEEEPARstart{P}{rinted} circuit board (PCB) design is traditionally performed using schematic capture and interactive routing within electronic design automation (EDA) tools.
This workflow typically follows three steps.
\begin{description}
	\item[\bf Step 1.] Capture the electronic schematic, including component symbols and their interconnections. \medskip

	\item[\bf Step 2.] Assign footprints to each symbol.  \medskip

	\item[\bf Step 3.] Place components and route copper tracks, adding other elements such as edge cuts, planes, vias, keepout zones, and graphical annotations.
\end{description}

In particle physics experiments, however, detector readout electronics often require precise, repetitive, and parameterized copper geometries to meet stringent impedance and signal integrity constraints.
For such applications, this conventional workflow becomes cumbersome, particularly when hundreds of similar structures (e.g., Resistive Plate Chamber (RPC) readout strips or differential signal pairs) must be generated or iterated upon.
For instance, the CMS RPC upgrade and the COMET Cosmic Ray Veto system demand co-optimized detector geometry and signal routing, where manual design is error-prone and time-consuming.
\kpcbn addresses this gap by enabling the fully programmatic generation of KiCad PCB files, effectively merging the three conventional steps into a single script-driven process without requiring a graphical interface.
\begin{figure}
    \centering
    \includegraphics[width=1.00\linewidth]{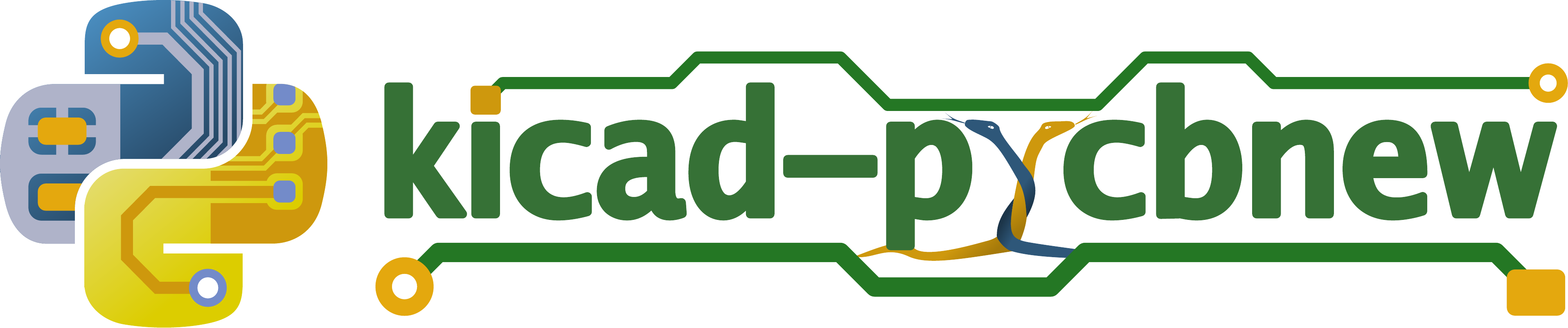}
		\caption{\kpcbn logo. The official Python logo adapted for \kpcbn. The traditional yellow and blue colors remain dominant, while PCB traces inspired by those visible on the \pcbnew logo are integrated into the design. The project name is rendered in the green of \pcbnew, with the letter “y” formed by two intertwined Python snakes (yellow and blue) to symbolize Python’s integration into \pcbnew.}
    \label{fig:logo-wordmark}
\end{figure}

\section{Background and motivation}
	\subsection{KiCad as a foundation for fully scripted PCB design}
Among available EDA tools, KiCad was chosen as the target platform for several reasons.
It is a free and open-source suite with documented, human-readable file formats, ensuring transparency, long-term accessibility of design data, and reproducibility of results.
Its wide adoption in both academia and industry, together with sustained development and institutional support ---~most notably from CERN\cite{del2015kicad,del2015kicadchallenges,raynova2017kicad}~--- makes it a reliable and sustainable foundation for scientific and engineering projects.

Within KiCad, the PCB layout stage (Step 3) is handled by the \pcbnew editor.
Although \kpcbn is closely connected to \pcbnew and its file format ---~as suggested by its logo (see logo Fig.~\ref{fig:logo-wordmark})~---, the project is not limited to scripting only the layout stage.
It enables a fully programmatic description of the board, effectively integrating and replacing the three conventional design steps.

By working directly with a structured Python object model rather than existing KiCad projects, \kpcbn allows users to define boards entirely in code.
This approach supports reproducible, parameterized, and automated PCB generation, while maintaining precise control over both high-level structures and low-level details written to the final KiCad board file.

	\subsection{Statement of need}

Printed circuit boards used in particle detectors such as Resistive Plate Chambers (RPCs) often require the precise definition of large copper geometries, including readout strips and controlled-impedance signal traces.
Such constraints arise in several detector electronics developments, for example, in the CMS RPC upgrade~\cite{SHCHABLO2020162139,Meola2020} and in the Cosmic Ray Veto system of the COMET experiment~\cite{comet2020comet}.
More broadly, similar readout architectures appear in many RPC-related developments where signal routing and detector geometry must be co-optimized~\cite{Novel_RPC_Designs_2021,bielewicz2023practical}.
Designing and iterating on these geometries with conventional interactive PCB tools can be time-consuming, especially when hundreds of similar structures must be generated or modified programmatically.

One may wonder why a new tool is necessary, given existing KiCad scripting capabilities.
The KiCad graphical interface already provides scripting capabilities within \pcbnew to automate repetitive layout operations.

\kpcbn and the KiCad Python API address related but distinct use cases in the KiCad ecosystem.
The official API \cite{KiCad_Python_API} provides Python bindings to KiCad’s internal C++ data structures and is primarily intended for extending or automating the PCB editor within the KiCad environment (step 3).
It enables direct manipulation of existing boards but remains tightly coupled to KiCad’s runtime.
In contrast, \kpcbn is a standalone Python library that generates \kicadpcb files directly from a structured object model, without requiring KiCad to be installed or running.
This approach favors deterministic, script-driven PCB generation and straightforward integration into headless workflows and CI/CD (Continuous Integration and Continuous Delivery) pipelines.
While it does not provide interactive access to KiCad’s editor or internal DRC engine, it offers a lightweight alternative for programmatic, parametric, or large-scale PCB generation.

Other tools such as
\begin{itemize}
	\item \texttt{kicad-skip}~\cite{kicad_skip},
	\item \texttt{KiCadFiles}~\cite{KiCadFiles},
	\item \texttt{kicad-pcb-api}~\cite{kicad_pcb_api},
	\item \texttt{kicad-parts-placer}~\cite{kicad_parts_placer},
	\item \texttt{PCBFlow}~\cite{pcbflow}
\end{itemize}
address file manipulation, automated placement, or script-based PCB generation within the conventional workflow.

In these tools, as well as in other programmatic hardware design approaches such as SKiDL~\cite{skidl}, the classical architecture remains unchanged:
(step 1) a schematic defining symbols, logic and nets,
(step 2) a symbol-to-footprint association stage,
and (step 3) a separate PCB layout and routing phase.

\kpcbn departs from this architecture by unifying these three steps into a single programmatic model.
The board is not derived from a prior schematic project but defined directly as code.
Electrical connectivity, footprint instantiation, component placement, and copper geometry are described within the same script and generated in a single deterministic process.

This integration removes the need for an intermediate schematic representation and enables parameterized, reproducible, and fully automated PCB generation from scratch.
Compared to the other tools mentioned above, \kpcbn uniquely provides a fully standalone, script-driven approach that combines schematic, footprint, and layout generation in one coherent Python API.

\medskip
The main strengths of \kpcbn are summarized below.
\begin{itemize}
	\item \kpcbn shifts the design paradigm from interactive editing of project files to the declarative specification of complete boards through high-level abstractions.
	\item For boards with hundreds of similar traces ---~where one or more parameters may vary systematically~--- \kpcbn can reduce design time from days to hours while ensuring geometric precision.
	\item Because it does not require a running KiCad instance, \kpcbn can be integrated into CI/CD workflows.
\end{itemize}

\section{Implementation and usage}

	\subsection{Architecture}

KiCad \kicadpcb\ files are structured as S-expressions, a human-readable nested text format describing the full PCB layout as a hierarchy of objects.
\kpcbn is implemented in Python ($\geq$3.10) and generates these files directly through structured text output fully compliant with the official KiCad file format specification.
\kpcbn constructs the complete board description in Python and serializes it directly into a valid \kicadpcb file, without using KiCad’s internal bindings or GUI.
Of course, KiCad remains essential for downstream tasks such as visualizing the board, running design rule checks (DRC), exporting manufacturing files (e.g., Gerber), and producing 3D renderings.

\begin{figure*}
    \centering
    \includegraphics[height=0.6\linewidth]{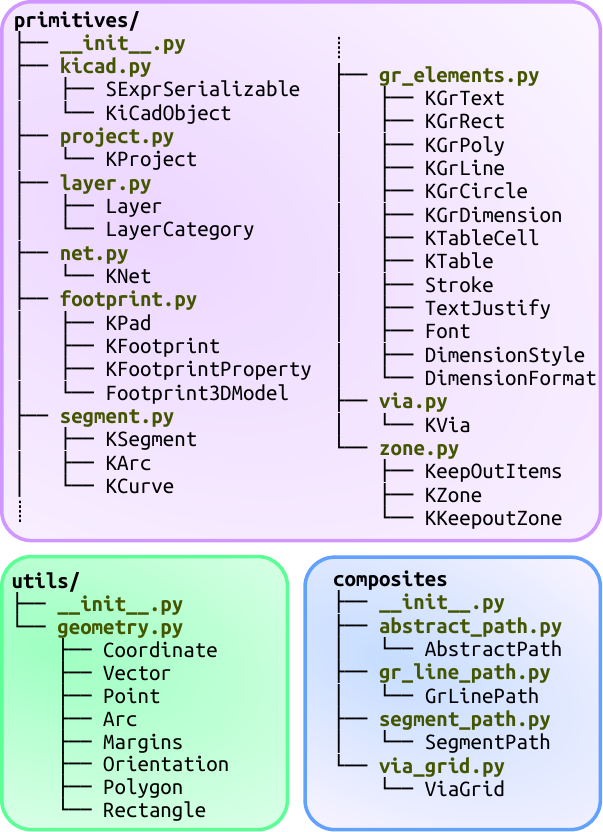}\qquad
	\includegraphics[height=0.6\linewidth]{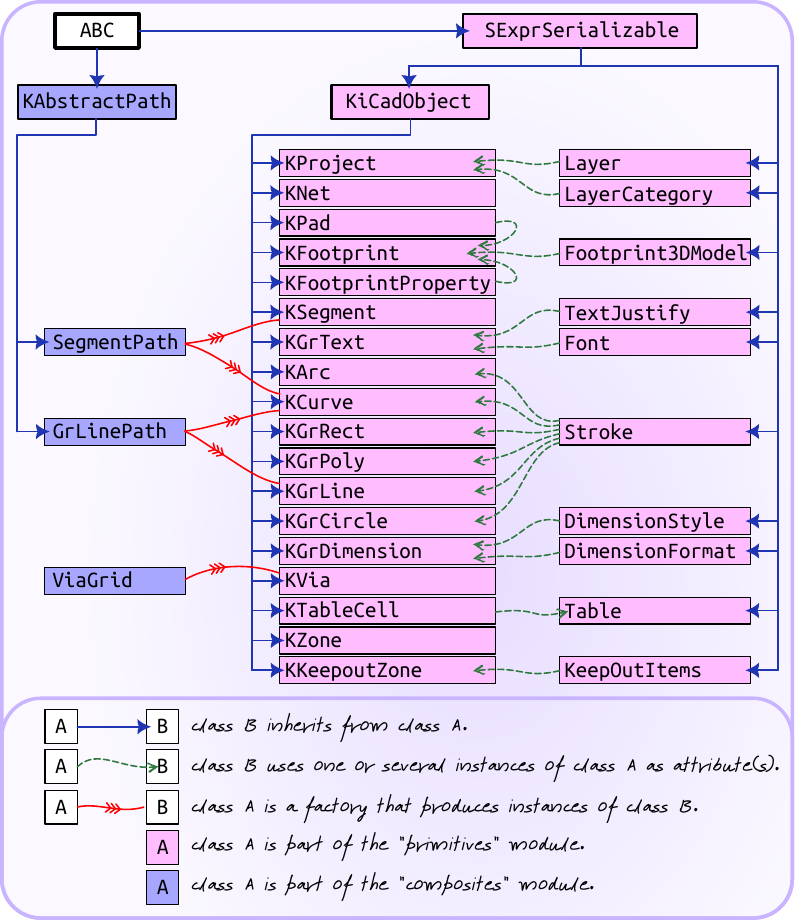}
    \caption{\textbf{Left:} directory and class structure of the \kpcbn{} project, showing the three modules, Python files, and the classes defined in each file. \textbf{Right:} Class hierarchy and relationships in \kpcbn. Most PCB primitives inherit from \texttt{KiCadObject}, itself derived from \texttt{SExprSerializable}, enabling direct serialization to KiCad’s S-expression format. Arrows indicate inheritance, composition, and factory relationships between core objects (e.g., \texttt{KGrLine} inherits from \texttt{KiCadObject}, uses \texttt{Stroke} instances as internal parameters, and is potentially produced by \texttt{GrLinePath}).}
    \label{fig:project_structure_and_hierarchy}
\end{figure*}

\medskip
The architecture is organized into three main Python modules with clearly separated responsibilities:
	a \textbf{geometry} module,
	a KiCad \textbf{primitives} module,
	and a \textbf{higher-level} layout module (see Fig.~\ref{fig:project_structure_and_hierarchy} for file structure (left) and class hierarchy (right)).
This layered architecture separates geometry manipulation, PCB object representation, and layout generation logic.
It allows users to describe complex PCB structures at a high level while maintaining precise control over the low-level elements written to the final KiCad board file.

Note that a complete description of the architecture and APIs is available in the project documentation hosted on Read the Docs (\href{https://kicad-pycbnew.readthedocs.io}{https://kicad-pycbnew.readthedocs.io}).

	\subsubsection{A general-purpose geometry module}
	The \texttt{geometry} module provides the mathematical building blocks used throughout the project.
	It defines fundamental geometric objects such as points, vectors, arcs, rectangles, and polygons, along with utilities for handling orientations and margins.
	Each class provides convenient methods that help developers maintain high-level code.
	For example, the module implements common geometric transformations including translations, rotations, polygon growth and shrinkage (offset operations, useful for defining margins or clearances), geometric computations (such as intersections and point-in-polygon problem solver) and polygon boolean operations such as subtraction.
	These tools allow complex PCB geometries to be constructed independently of the KiCad file representation.

	Importantly, the geometry module accounts for KiCad’s coordinate convention, where the Y-axis points downward while maintaining the usual counterclockwise (trigonometric) orientation, resulting in an inverted base. Example:
	\[\texttt{Vector(1, 0)} \xrightarrow{\ \text{rotate}(\pi/2)\ } \texttt{Vector(0, -1)}\]

	\subsubsection{A KiCad primitives module}
	The \texttt{primitives} implements the core PCB data model corresponding to the atomic elements present in KiCad boards.
	This includes objects representing projects, layers, nets, footprints and pads, track segments and arcs, vias, copper zones, and graphical elements (texts, lines, polygons, circles, tables, dimensions).

	Each of these objects is responsible for its serialization into the KiCad S-expression format, thereby acting as a bridge between Python objects and the textual PCB representation used by KiCad.
	Thus, all primitive PCB elements share a common serialization mechanism based on two abstract base classes.
	\begin{enumerate}
		\item At the foundation, \texttt{SExprSerializable} defines the generic interface for any object that can be represented as a KiCad S-expression.
	Subclasses implement the \texttt{sexp\_tree()} method, which returns a nested list describing the hierarchical structure of the corresponding S-expression, later converted into the textual \texttt{.kicad\_pcb} format.
	Each object therefore both defines its S-expression structure, providing a consistent and extensible representation of PCB elements.

		\item Building on this interface, the class \texttt{KiCadObject} introduces functionality specific to KiCad board elements by automatically assigning a universally unique identifier (UUID) to each instance, corresponding to the identifiers used internally by KiCad.
	Most primitive layout classes (e.g. \texttt{KSegment}, \texttt{KVia}, \texttt{KZone}, \texttt{KFootprint}\ldots) inherit from this class and follow a naming convention in which their names start with the letter \texttt{K}.
	\end{enumerate}

	\subsubsection{A high-level layout module}
	The \texttt{composites} module provides convenient layout utilities that build complex PCB structures from combinations of primitives.
	These abstractions simplify the generation of repetitive or geometrically constrained layouts.
	Examples include:
	\begin{itemize}
		\item multi-segment tracks represented as parametric paths;
		\item curved track segments or graphical lines where sharp corners can
		be automatically replaced by Bézier curves;
		\item sets of parallel equidistant tracks derived from a reference path;
		\item track width transitions used for controlled impedance changes;
		\item structured grids of vias and polygon-based exclusion zones.
	\end{itemize}

	\subsection{Workflow}
\begin{figure}
    \centering
    \includegraphics[width=1.00\linewidth]{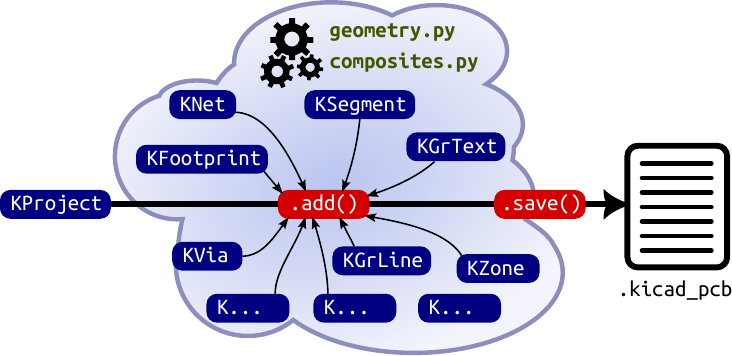}
    \caption{Overview of the \kpcbn workflow: project initialization with \texttt{KProject}, creation and positioning of primitives with the \texttt{geometry }module, addition to the project, and final PCB file generation through the \texttt{.add()} method.}
    \label{fig:workflow}
\end{figure}

The overall workflow of a \kpcbn project is illustrated in Fig.~\ref{fig:workflow}.
A \texttt{KProject} object is first created, geometric primitives are positioned and managed using the \texttt{geometry} and \texttt{composites} modules before being added via the \texttt{.add()} method.
The \texttt{.save()} method finally generates and writes the corresponding \kicadpcb file in S-expression format, which can be opened and analysed with KiCad \texttt{pcbnew}.

	\subsection{Minimal working example}
To illustrate the simplicity of creating a new PCB project and adding basic elements, the code snippet below demonstrates how to instantiate a \texttt{KProject}, define a ground net, place a resistor footprint, and save the result.

\begin{lstlisting}[language=Python,backgroundcolor=\color{white}]
from pycbnew.primitives.project import KProject
from pycbnew.primitives.footprint import KFootprint
from pycbnew.primitives.net import KNet
from pycbnew.utils.geometry import Point

# Create a new KiCad project with 2 copper layers
project = KProject(filename="myBoard.kicad_pcb",
                   copper_layers=2)

# Define a ground net and add it to the project
gnd_net = KNet(1, "GND")
project.add(gnd_net)

# Add a resistor footprint to the front side
footprint = KFootprint(
    "Resistor_SMD:R_0805_2012Metric",
    at=Point(100, 100),
    side=KFootprint.Side.Front,
    default_net=gnd_net
)
project.add(footprint)

# Save the project to a .kicad_pcb file
project.save()
\end{lstlisting}

This example highlights the declarative nature of the library: users define the board structure programmatically, and the final KiCad-compatible file is generated in a single step.
The modular design of the \texttt{KProject}, \texttt{KNet}, and \texttt{KFootprint} classes ensures that complex layouts can be constructed by combining these primitives, while the \texttt{Point} utility from the geometry module provides precise control over component placement.

\medskip
Currently, \kpcbn generates PCB files compatible with the KiCad 9 file format.
Newer versions of KiCad can open these files thanks to the backward compatibility of the format, while older versions may not.
Future releases of \kpcbn will evolve alongside KiCad to support newer file formats and design capabilities as they become available.

\section{Documentation and evaluation}

	\subsection{Example scripts and tutorials}
The project, hosted on \href{https://gitlab.com/vinraspa/kicad-pycbnew}{Gitlab}, comes with multiple scripts (`\texttt{examples/}' folder) of increasing complexity intended to demonstrate board generation scenarios, including trapezoidal RPC readout boards and extension boards with merged impedance-controlled traces.
These examples serve both as documentation, tutorials, and reproducible test cases.
A brief overview of the example scripts is provided in Table~\ref{tab:example_scripts}, separating simpler tutorial scripts from full-scale realistic PCB projects, illustrating both learning workflows and real-world applications.
\begin{table}
	\centering
	\caption{Example scripts for \kpcbn{}.
	Examples 1–8 (tutorial-style scripts) illustrate basic operations and step-by-step usage.
	Examples 9–10 (full-scale scripts) demonstrate realistic PCB generation for RPC readout and extension boards
	All scripts are available on \href{https://gitlab.com/vinraspa/kicad-pycbnew/-/tree/main/examples}{GitLab}.}
	\label{tab:example_scripts}
	\begin{tabular}{p{2.5cm}p{5.5cm}}
		\toprule
		\textbf{Example file} & \textbf{Purpose and features}\\
		\midrule
		\fnametab{01\usc{}project.py} &
		Creates a basic KiCad PCB project: customizes copper layers, paper formats, and saves the project. \\
		\midrule
		\fnametab{02\usc{}nets.py} &
		Creates and manages nets (creation and addition to a project). \\
		\midrule
		\fnametab{03\usc{}footprints.py} &
		Works with footprints: path specification (absolute, relative, KiCad format), placement, rotation, board side, net assignment to pads, and modifications of properties (reference, value, etc.). \\
		\midrule
		\fnametab{04\usc{}traces.py} &
		Creates PCB traces: individual straight traces (\texttt{KSegment}), chained segments (\texttt{KSegmentPath}), quick trace path creation, and multi-layer copper trace management. \\
		\midrule
		\fnametab{05\usc{}traces2.py} &
		Shows advanced trace techniques: complex paths between footprints, parallel traces, angular and curved paths (Fig.~\ref{fig:parallel_curved_tracks}). \\
		\midrule
		\fnametab{06\usc{}zones.py} &
		Introduces copper zones and keepout areas: geometric zone creation, multi-layer application, fill order control, keepout zones, and grid generation (excluding keepout polygons). \\
		\midrule
		\fnametab{07\usc{}vias.py} &
		Focuses on vias: through-hole, blind, or buried, custom dimensions, and via grids for stitching/thermal purposes. \\
		\midrule
		\fnametab{08\usc{}graphics\usc{}and\usc{}edge\usc{}cuts.py} &
		Draws graphical items and defines board outlines: board edge cuts, regular polygons, polygon resizing, centered/multi-line labels, text justification, tables, and linear dimensions. \\
		\midrule
		\fnametab{09\usc{}RPC\usc{}readout\usc{}board.py} &
		Demonstrates an advanced use of \kpcbn by creating a trapezoidal iRPC readout board, similar to those used in the CMS experiment (see \emph{Figure~4} in \cite{SHCHABLO2020162139}).
		The script is fully functional, and users can customize parameters to adapt the design.
		Preview the board \href{https://gitlab.com/vinraspa/kicad-pycbnew/-/blob/main/examples/09_RPC_readout_board.svg}{here} (Fig.~\ref{fig:RPC_readout_PCB}).\\
		\midrule
		\fnametab{10\usc{}RPC\usc{}extension\usc{}board.py} &
		This script uses \kpcbn to create an extension board between an iRPC readout board and a FEB, increasing distance to cut noise and radiation.
		It pairs strip signals, reducing FEB channel needs during low-intensity readout.
		The example also covers ground via grids with exclusion zones and regions for logarithmic impedance transitions.
		Preview the board \href{https://gitlab.com/vinraspa/kicad-pycbnew/-/blob/main/examples/10_RPC_extension_board.svg}{here} (Fig.~\ref{fig:RPC_extension_board}).\\
		\bottomrule
	\end{tabular}
\end{table}

\begin{figure}
	\centering
	\includegraphics[width=0.95\linewidth]{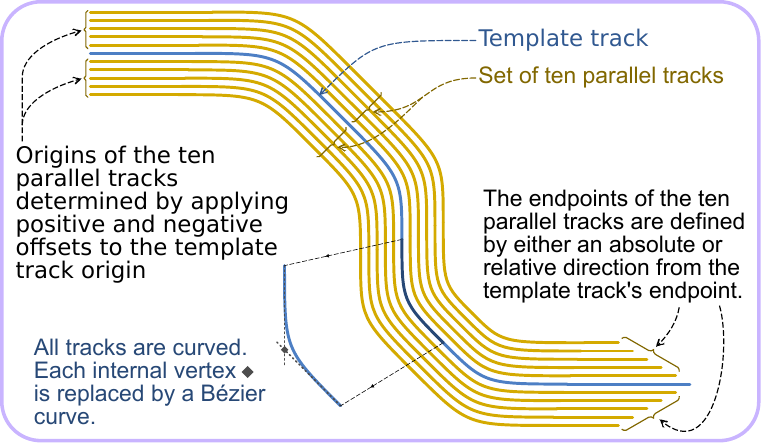}
	\caption{Example of parallel curved tracks generated using the script \texttt{05\_traces2.py}. The central track is used as a template to generate ten additional parallel traces. The origin and endpoints of the parallel tracks are determined according to user-defined rules. The tracks are curved: all internal vertices are replaced by Bézier curves that ensure smooth transitions and minimize signal reflections ---~critical for maintaining signal integrity when traces support high-speed signals.}
	\label{fig:parallel_curved_tracks}
\end{figure}

\begin{figure*}
	\centering
	\includegraphics[width=0.75\linewidth]{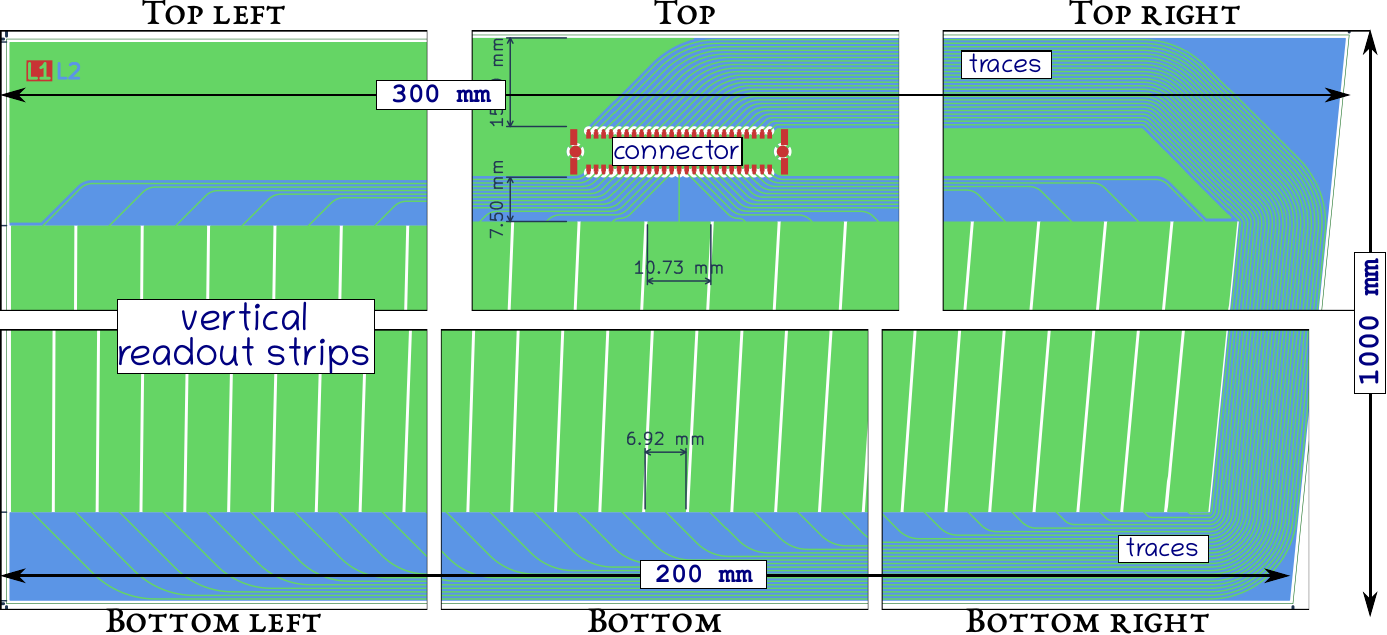}
	\caption{Sections of a $1000\times300~\rm mm^2$ RPC readout board generated with the script \texttt{09\_RPC\_readout\_board.py}.
The figure shows the strip layout, the returning parallel traces and their connections to footprint pads, the via transitions, and the inclusion of user-defined annotations such as dimensions.
A complete view of the board is available in the project's GitLab repository.}
	\label{fig:RPC_readout_PCB}
\end{figure*}

\begin{figure}
	\centering
	\includegraphics[height=0.65\linewidth,angle=90]{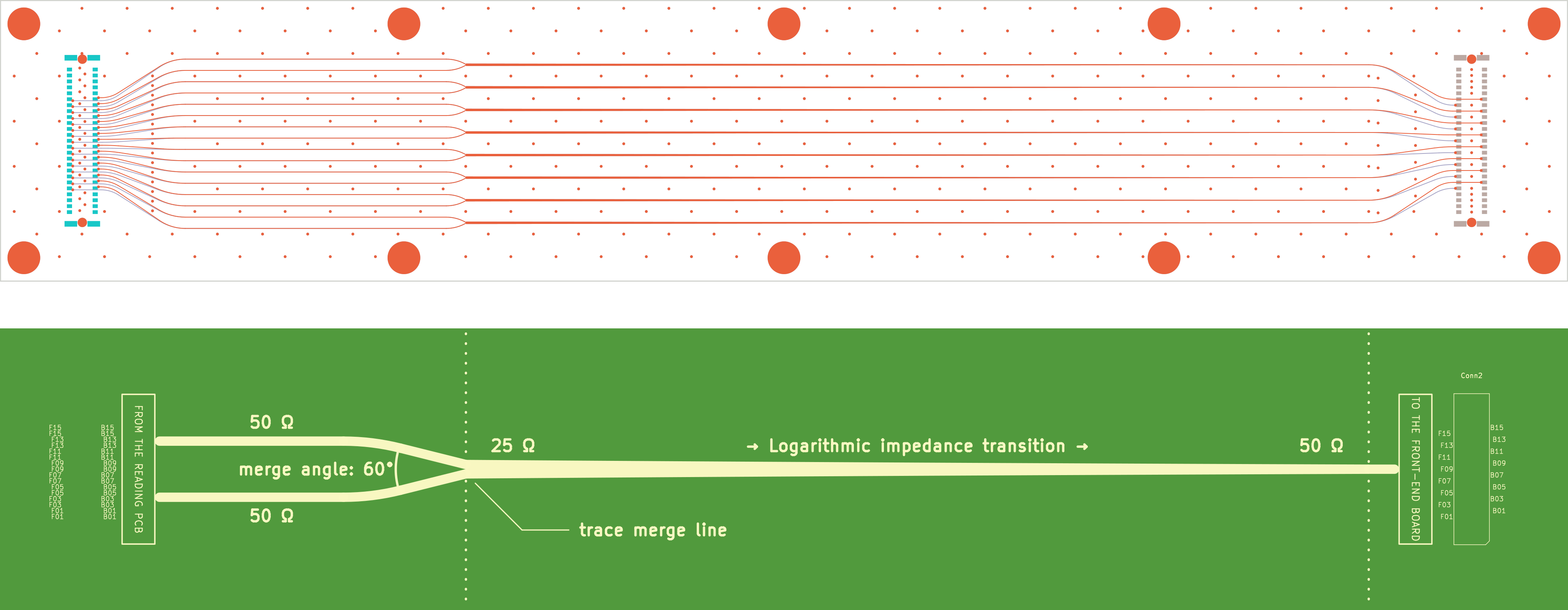}
	\caption{Inner copper and top silkscreen layers of an extension board designed with \kpcbn (script \texttt{10\_RPC\_extension\_board.py}) for \href{https://comet.kek.jp/}{COMET project} to be placed between an iRPC readout board and a Front-End Board (FEB), allowing the FEB to be positioned farther away to reduce noise and radiation exposure. The figure shows traces emerging from a connector, diverging, and then merging in pairs, finally passing through a logarithmic impedance transition between traces of different widths. The traces are routed on a structured grid with multiple ground vias, including user-defined exclusion zones (\kpcbn feature) to maintain signal integrity.}
	\label{fig:RPC_extension_board}
\end{figure}

	\subsection{Availability}

The software is compatible with Linux, Windows, and macOS systems, requiring only an environment capable of running Python 3.10 or higher and KiCad version 9 or later.
It is implemented in Python ($\geq$ 3.10) and does not impose specific hardware constraints beyond those necessary to run Python and KiCad; memory and disk usage remain modest and scale with the complexity of the generated boards.
The software relies on a minimal set of dependencies, namely the Python modules \texttt{strenum} and \texttt{pyparsing}, the latter being used for parsing footprint files.
An archived version of the software is available on \href{https://doi.org/10.5281/zenodo.18924130}{Zenodo}, published under the MIT licence (version 1.0.1, 2026/02/24).
The source code is maintained in a \href{https://gitlab.com/vinraspa/kicad-pycbnew}{public Gitlab repository}, also distributed under the MIT licence, with initial publication dating from 2024/11/29.

Installation can be performed with \texttt{pip}, ideally inside a virtual environment.
\begin{lstlisting}[backgroundcolor=\color{white}]
$ python -m venv ./venv/
$ source ./venv/bin/activate
$ pip install kicad-pycbnew
\end{lstlisting}

	\subsection{Quality control}

The reliability of the software is ensured through several complementary mechanisms combining automated testing and validation through practical use.

\begin{itemize}
	\item Standard Python unit tests are implemented with the \texttt{unittest} framework.
	These tests cover the core classes and methods of the library and verify the correct behavior of geometric primitives, PCB objects, and layout utilities.
	Both the internal Python objects and the generated KiCad S-expressions are checked to ensure that the resulting structures conform to the expected file format.

	\item A collection of structured example scripts that generate complete and	non-trivial boards.
	These examples exercise typical workflows and act as integration-level tests for the high-level API.
	Their good execution is also part of the tests described above.

	\item Validation of generated boards by opening them in KiCad, ensuring that
	the produced \kicadpcb\ files are correctly parsed and fully compatible with
	the official file format.

	\item Iterative use of the library in real detector PCB development workflows,
	providing continuous feedback on robustness and practical usability.

	\item Manual inspection and design rule checking (DRC) within KiCad on the
	generated boards.
\end{itemize}

Users can verify a correct installation by running one of the provided example scripts and opening the generated \kicadpcb\ file in KiCad.
Successful rendering of the board without file parsing errors confirms that the library operates correctly.

The project repository also includes documentation and usage examples that allow new users to quickly evaluate the capabilities of the software.
The library has already been used to generate printed circuit boards for RPC detector readout applications.
These boards have been manufactured and experimentally tested, while detailed performance measurements will be reported in future dedicated publications.

	\subsection{Reuse potential}
\kpcbn can be reused in any domain where programmable PCB layout is beneficial.
Potential reuse scenarios include:

\begin{itemize}
	\item detector readout boards in particle physics experiments;
	\item high-speed or impedance-controlled PCB designs;
	\item automated generation of repetitive or patterned copper geometries;
	\item parametric board families derived from shared geometric constraints;
	\item educational contexts for demonstrating PCB file structure and automated design.
\end{itemize}

The modular architecture allows extension through additional geometric primitives in \texttt{geometry} module (e.g. introducing new KiCad primitives as the software evolves) or higher-level layout functions (new primitive factories).
Researchers can fork the public repository and contribute via merge requests.
Documentation and example scripts lower the barrier to adoption.
Support is provided through the public GitLab issue tracker associated with the repository.
\kpcbn can also be integrated into automated pipelines or batch-generation workflows (e.g., CI/CD) for large-scale or parametric PCB production, enhancing reproducibility and efficiency.

\section{Conclusion}
\kpcbn provides a programmable approach to PCB generation by directly producing KiCad-compatible board files from a structured Python API. By decoupling board definition from interactive CAD environments, it enables deterministic and reproducible design workflows, particularly suited to parameterized and geometry-driven layouts.
The abstraction layers implemented in the library allow users to construct complex copper structures, including parallel trace bundles, impedance-controlled transitions, and via networks, with a concise and maintainable codebase.
The software has been validated through the generation of full-scale detector readout boards, demonstrating its capability to handle realistic design constraints and large geometries.
Its modular architecture facilitates extension through additional geometric primitives and layout strategies, making it adaptable to a wide range of PCB design problems.
As such, \kpcbn complements existing EDA tools by providing a flexible, script-based alternative for automated PCB generation.

\section*{Acknowledgments}
This work was supported by the COMET collaboration, Laboratoire de Physique Clermont Auvergne and Université Clermont Auvergne.

\bibliographystyle{IEEEtran}
\bibliography{arXiv_references}

@misc{KiCad_Python_API,
  author       = {{The KiCad Development Team }},
  title        = {KiCad Python API},
  howpublished = {\url{https://gitlab.com/kicad/code/kicad-python/}},
  year         = {2025}
}

@misc{kicad_skip,
  author       = {Deegan, Pat},
  title        = {kicad-skip},
  howpublished = {\url{https://github.com/psychogenic/kicad-skip}},
  year         = {2023}
}

@misc{KiCadFiles,
  author       = {Steffen, W.},
  title        = {KiCadFiles},
  howpublished = {\url{https://steffen-w.github.io/KiCadFiles/}},
  year         = {2024}
}

@misc{kicad_pcb_api,
  author       = {{circuit-synth}},
  title        = {kicad-pcb-api},
  howpublished = {\url{https://pypi.org/project/kicad-pcb-api/}},
  year         = {2023}
}

@misc{kicad_parts_placer,
  author       = {Hobbs, Simon},
  title        = {kicad-parts-placer},
  howpublished = {\url{https://pypi.org/project/kicad-parts-placer/}},
  year         = {2023}
}

@misc{del2015kicad,
  title={KiCad software gets the CERN treatment},
  author={Del Rosso, Antonella},
  howpublished = {\url{https://home.cern/news/news/computing/kicad-software-gets-cern-treatment}},
  year={2015}
}

@misc{del2015kicadchallenges,
  title={KiCad challenges the big ones},
  author={Del Rosso, Antonella},
  howpublished = {\url{https://repository.cern/records/bagx4-a9560}},
  year={2015}
}

@misc{raynova2017kicad,
  title={KiCad reaches new heights},
  author={Raynova, Iva},
  howpublished = {\url{https://cds.cern.ch/record/2276570}},
  year={2017}
}

@article{Meola2020,
  title={Towards a two-dimensional readout of the improved CMS Resistive Plate Chamber},
  author={Meola, Sabino et al.},
  journal={arXiv preprint arXiv:2006.00576},
  year={2020}
}

@article{comet2020comet,
  title={COMET Phase-{I} technical design report},
  author={{Comet Collaboration}},
  journal={Progress of Theoretical and Experimental Physics},
  volume={2020},
  number={3},
  pages={033C01},
  year={2020},
  publisher={Oxford University Press}
}

@article{Novel_RPC_Designs_2021,
  title={Development of Novel Designs of Resistive Plate Chambers},
  author={{Calice Collaboration}},
  journal={Instruments},
  year={2021},
  url = {https://cds.cern.ch/record/2861342/files/document.pdf},
  publisher={MDPI}
}

@article{bielewicz2023practical,
  title={Practical Implementation of an Analogue and Digital Electronics System for a Modular Cosmic Ray Detector—MCORD},
  author={Bielewicz, Marcin et al.},
  journal={Electronics},
  volume={12},
  number={6},
  pages={1492},
  year={2023},
  publisher={MDPI},
  url={https://www.mdpi.com/2079-9292/12/6/1492}
}

@article{SHCHABLO2020162139,
title = {Performance of the CMS RPC upgrade using 2D fast timing readout system},
author = {K. Shchablo and I. Laktineh and M. Gouzevitch and C. Combaret and L. Mirabito},
journal = {Nuclear Instruments and Methods in Physics Research Section A: Accelerators, Spectrometers, Detectors and Associated Equipment},
volume = {958},
pages = {162139},
year = {2020},
note = {Proceedings of the Vienna Conference on Instrumentation 2019},
issn = {0168-9002},
doi = {10.1016/j.nima.2019.04.093},
url = {https://www.sciencedirect.com/science/article/pii/S0168900219305820},
}

@misc{skidl,
  author       = {Devbisme},
  title        = {SKiDL: A Python-Based Schematic Design Language},
  howpublished = {\url{https://github.com/devbisme/skidl}},
  year         = {2023}
}

@misc{pcbflow,
  author       = {Gale, Michael},
  title        = {PCBFlow},
  howpublished = {\url{https://github.com/michaelgale/pcbflow}},
  year         = {2022}
}

\end{document}